\documentclass[twocolumn,showpacs,aps,prl,superscriptaddress,preprintnumbers,floatfix]{revtex4}

\usepackage{graphicx}
\usepackage{subfigure}
\usepackage{dcolumn}
\usepackage{bm}

\newcommand{\BABARPubYear}    {04}
\newcommand{\BABARPubNumber}  {041}
\newcommand{\SLACPubNumber} {11009}

\usepackage{relsize}
\usepackage{xspace}
\def\babar{\mbox{\slshape B\kern-0.1em{\smaller A}\kern-0.1em
    B\kern-0.1em{\smaller A\kern-0.2em R}}}
\def\pep2{PEP-II}
\def\CP{\ensuremath{C\!P}\xspace}

\def\pipm  {\ensuremath{\pi^\pm}\xspace}

\def\Kbar  {\kern 0.2em\overline{\kern -0.2em K}{}\xspace}

\def\Kstarzb {\ensuremath{\Kbar^{*0}}\xspace}

\def\Dp      {\ensuremath{D^+}\xspace}

\def\Dpm     {\ensuremath{D^\pm}\xspace}

\def\Ds      {\ensuremath{D^+_s}\xspace}
\def\mphi       {\mbox{$\phi$}\xspace}
\def\kkpi       {\ensuremath{K^{-}K^{+}\pi^{+}}\xspace}
\def\Dptokkpi   {\ensuremath{\Dp \to K^{-}K^{+}\pi^{+}}\xspace}

\def\Dpmtokkpi  {\ensuremath{\Dpm \to K^{-}K^{+}\pi^{\pm}}\xspace}
\def\Dpmtophipi {\ensuremath{\Dpm \to \phi\pi^{\pm}}\xspace}
\def\Dpmtokstark {\ensuremath{\Dpm \to K^{*0}K^{\pm}}\xspace}
\def\Dptokpipi  {\ensuremath{\Dp \to K^{-}\pi^{+}\pi^{+}}\xspace}

\def\Dpmtokpipi {\ensuremath{\Dpm \to K^{\mp}\pi^{\pm}\pi^{\pm}}\xspace}
\def\Dptophipi  {\ensuremath{\Dp \to \mphi\pi^{+}}\xspace}
\def\Dptokstark {\ensuremath{\Dp \to \Kstarzb K^{+}}\xspace}

\def\Dsptokkpi  {\ensuremath{\Ds \to K^{-}K^{+}\pi^{+}}\xspace}
\def\Dspmtokkpi {\ensuremath{D^{\pm}_{s} \to K^{-}K^{+}\pi^{\pm}}\xspace}
\newcommand{\gevc}{\ensuremath{{\mathrm{\,Ge\kern -0.1em V\!/}c}}\xspace}
\newcommand{\mevc}{\ensuremath{{\mathrm{\,Me\kern -0.1em V\!/}c}}\xspace}
\newcommand{\gevcc}{\ensuremath{{\mathrm{\,Ge\kern -0.1em V\!/}c^2}}\xspace}
\newcommand{\mevcc}{\ensuremath{{\mathrm{\,Me\kern -0.1em V\!/}c^2}}\xspace}
\newcommand{\dedx}{\ensuremath{dE/dx}\xspace}
\def\invfb   {\ensuremath{\mbox{\,fb}^{-1}}\xspace}

\begin{document}

\begin{flushleft}
\small{
\babar-PUB-\BABARPubYear/\BABARPubNumber \\
SLAC-PUB-\SLACPubNumber \\
}
\end{flushleft}
\vskip 0.5cm

\title{\boldmath 
A search for $C\!P$ violation and a measurement of the relative 
branching fraction in \Dptokkpi decays
}

%
\author{B.~Aubert}
\author{R.~Barate}
\author{D.~Boutigny}
\author{F.~Couderc}
\author{Y.~Karyotakis}
\author{J.~P.~Lees}
\author{V.~Poireau}
\author{V.~Tisserand}
\author{A.~Zghiche}
\affiliation{Laboratoire de Physique des Particules, F-74941 Annecy-le-Vieux, France }
\author{E.~Grauges-Pous}
\affiliation{Universitad Autonoma de Barcelona, E-08193 Bellaterra, Barcelona, Spain }
% ...
\collaboration{The \babar\ Collaboration}
\noaffiliation

\date{\today}

\begin{abstract}
We report on a search for the \CP asymmetry in the singly
Cabibbo-suppressed decays \Dptokkpi
and in the resonant decays \Dptophipi and \Dptokstark based on a
data sample of $79.9$~\invfb recorded by the \babar\ detector.
We use the Cabibbo-favored \Dsptokkpi branching fraction as normalization  
in the measurements to reduce systematic uncertainties. 
The \CP asymmetries obtained are
$A_{CP}(K^{-}K^{+}\pipm) = 
  (1.4 \pm 1.0 (\textrm{stat.}) \pm 0.8 (\textrm{syst.}))\times 10^{-2}$,
$A_{CP}(\mphi\pipm) = 
  (0.2 \pm 1.5 (\textrm{stat.}) \pm 0.6 (\textrm{syst.}))\times 10^{-2}$, and
$A_{CP}(\Kstarzb K^{\pm}) = 
  (0.9 \pm 1.7 (\textrm{stat.}) \pm 0.7 (\textrm{syst.}))\times 10^{-2}$.
The relative branching fraction  
$\frac{\Gamma (\Dptokkpi )}{\Gamma(\Dptokpipi)}$ is also measured and
is found to be 
 $(10.7 \pm 0.1 (\textrm{stat.}) \pm 0.2 (\textrm{syst.}))\times 10^{-2}$.
\end{abstract}


\maketitle

\section{INTRODUCTION}

\begin{figure}
  \centering
  \subfigure[]{\includegraphics[width=.47\columnwidth]{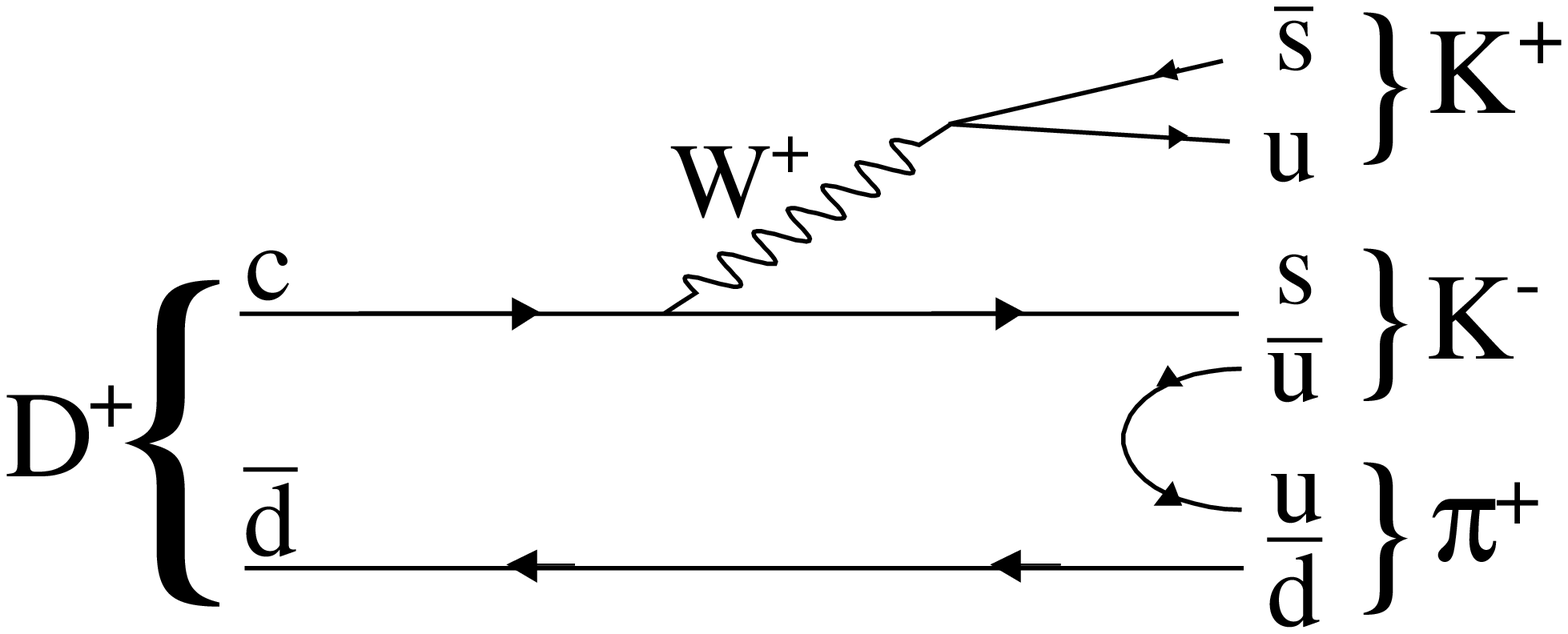}}
  \subfigure[]{\includegraphics[width=.47\columnwidth]{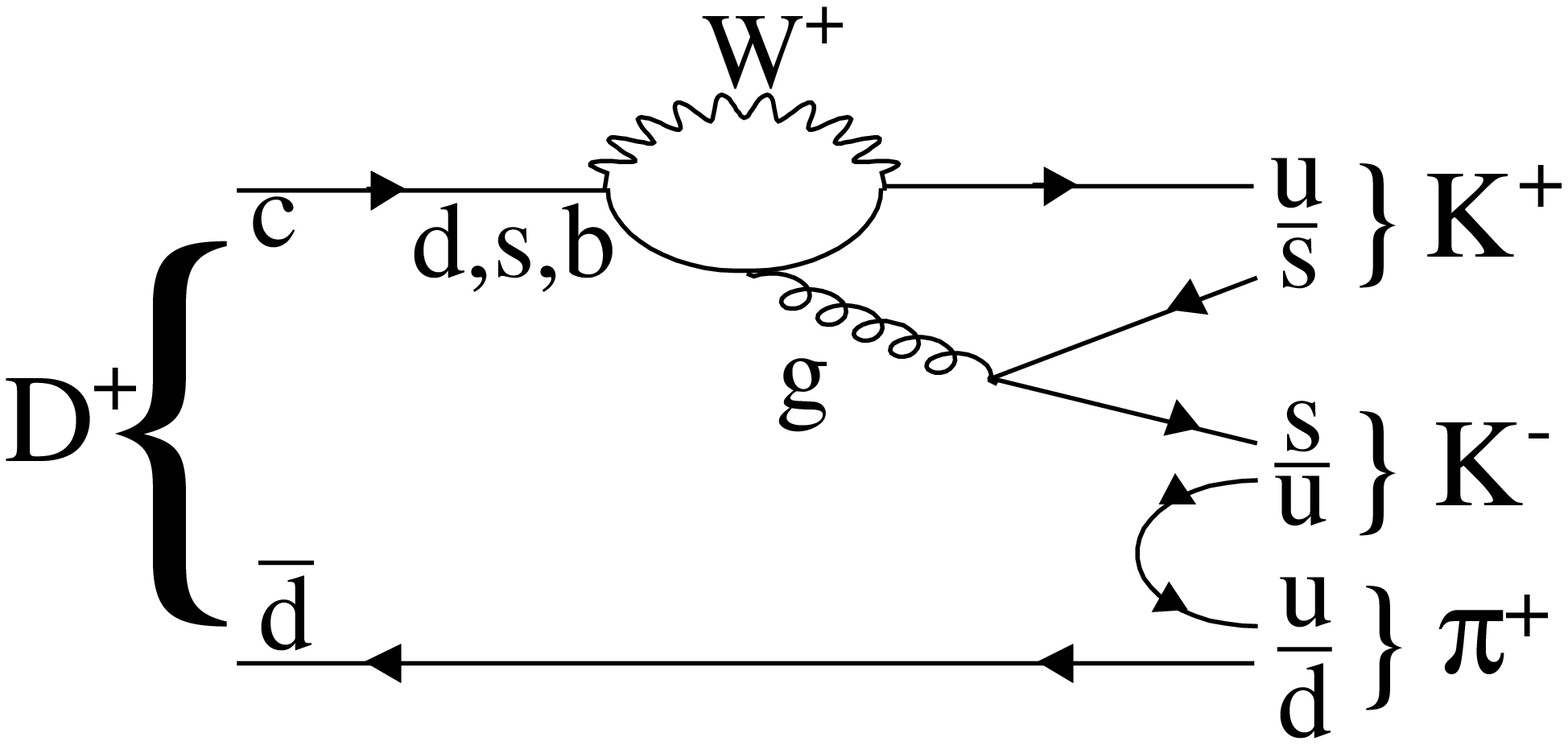}}
    \caption{\label{fig:Feynman}
      Parton-level diagrams for \Dptokkpi decays: 
      (a) a tree diagram, and (b) a penguin process.
    }
\end{figure}

Singly Cabibbo-suppressed (SCS) $D$-meson decays are predicted 
in the standard model (SM) to exhibit \CP-violating charge asymmetries of the 
order of $10^{-3}$~\cite{Buccella:1993}.
Direct \CP violation in SCS decays could arise from the 
interference between tree-level (Fig.~\ref{fig:Feynman}a) and penguin 
(Fig.~\ref{fig:Feynman}b) decay processes. Doubly Cabibbo-suppressed and 
Cabibbo-favored (CF) decays are expected to be \CP invariant in the SM because they 
are dominated by a single weak amplitude. Measurements of \CP asymmetries in SCS processes greater 
than $\cal O$$(10^{-3})$ would be evidence of physics beyond the 
standard model~\cite{Bianco}. 

We define the \CP asymmetry by 
\begin{equation}
A_{CP} =
\frac{\left|\mathcal{A}\right|^{2}-\left|\overline{\mathcal{A}}\right|^{2}}
     {\left|\mathcal{A}\right|^{2}+\left|\overline{\mathcal{A}}\right|^{2}},
\label{eq:one}
\end{equation}

\noindent where $\mathcal{A}$ is the total decay amplitude for \Dp
decays and $\overline{\mathcal{A}}$ is the amplitude for the
charge-conjugate decays. $A_{CP}$ is non-zero only if there are at least
two different decay amplitudes with a \CP-violating relative weak
phase and a \CP-conserving relative strong phase due to
final-state interactions.  Eq.~(\ref{eq:one}) can be expressed as an
asymmetry of branching fractions.
We assume that the total decay
rates for $D^+$ and $D^-$ are equal (CPT invariance).
Assuming further that CF decays are invariant under
\CP, we use branching fractions for CF decays as normalization factors
to reduce experimental systematics due to particle identification (PID)
and tracking: 

\begin{equation}
A_{CP} = 
 \frac{
   \frac{\mathcal{B}(D^+\rightarrow K^+K^-\pi^+)}
        {\mathcal{B}(D_s^+\rightarrow K^+K^-\pi^+)}
  -\frac{\mathcal{B}(D^-\rightarrow K^+K^-\pi^-)}
        {\mathcal{B}(D_s^-\rightarrow K^+K^-\pi^-)}
 }
 { \frac{\mathcal{B}(D^+\rightarrow K^+K^-\pi^+)}
        {\mathcal{B}(D_s^+\rightarrow K^+K^-\pi^+)}
  +\frac{\mathcal{B}(D^-\rightarrow K^+K^-\pi^-)}
        {\mathcal{B}(D_s^-\rightarrow K^+K^-\pi^-)}
 }.
\label{eqacp}
\end{equation}

(Throughout this paper we assume that the production of $D^+$ and
$D_s^+$ mesons is charge symmetric.)

We also measure the \CP asymmetry in the resonant decays \Dptophipi and
\Dptokstark, and determine 
the relative branching fraction  
$\frac{\Gamma (\Dptokkpi )}
      {\Gamma(\Dptokpipi)}$.

\section{DETECTOR AND DATA SAMPLE}

This analysis is performed with a data sample recorded on and below the
$\Upsilon (4S)$ resonance with the \babar\ detector at the \pep2
asymmetric-energy $e^+e^-$ storage rings at the Stanford Linear
Accelerator Center.

The \babar\ detector is
described in detail in Ref.~\cite{Babar}. The silicon vertex tracker
(SVT) and the 40-layer cylindrical drift chamber (DCH) embedded in a 1.5-T
solenoid measure the momenta and energy loss (\dedx) of charged
particles. A ring-imaging Cherenkov detector (DIRC) is used for
charged-particle identification. Photons are detected and electrons
identified with a CsI(Tl) electromagnetic calorimeter (EMC).

We split the 89.7~\invfb data sample into a randomly selected
subsample of 9.8~\invfb to optimize the selection criteria 
and the remainder (a 79.9~\invfb sample) for the final analysis. 
This procedure eliminates selection bias. We apply 
the same selection criteria to the CF and SCS modes whenever possible. 
We determine selection efficiencies from a sample (145~\invfb equivalent) of 
Monte Carlo (MC)~\cite{G4} generated $e^+e^- \to c\overline{c}$ events.

\section{DATA ANALYSIS}

We reconstruct \Dp and \Ds~\cite{CCsign} decays by selecting events
containing at least three charged tracks.  Tracks are required to have
at least 12 measured DCH coordinates, a minimum transverse momentum of
0.1~\gevc, and to originate within 1.5~cm in $xy$ (transverse to the
beam) and $\pm 10$~cm along the $z$-axis (along the $e^-$ beam) of the
nominal interaction point. Kaons are identified by a selection on the
ratio of likelihood functions derived from 
\dedx in the SVT and DCH, and from the Cherenkov angle and number of
photons in the DIRC. Pions are
identified as tracks that fail a loose kaon identification
criterion. The three charged tracks are further constrained to originate
at a common vertex, the fit for which is accepted if the $\chi^2$
satisfies $P(\chi^2)>1\%$. We 
reject \Dp and \Ds mesons from $B$ decays, and thereby reduce
backgrounds, by requiring that their 
momenta in the center-of-mass (CM) frame be above 2.4 \gevc.

In order to reduce the remaining combinatorial background we consider
likelihood ratios formed from the probability density functions (PDFs) of
the following discriminating variables for the \Dp and \Ds decays: CM
momentum ($p_{\mathrm{CM}}$), vertex-fit probability 
with a beam-spot constraint ($P_{\mathrm{BS}}(\chi^2)$), and the distance
in the $xy$-plane from the 
interaction point to the \Dp or \Ds vertex ($d_{xy}$). The quantity
$P_{\mathrm{BS}}(\chi^2)$ is the probability that the decay tracks form
a vertex within the beam spot region. Most of the $D^+$ mesons decay
outside this region, thus the probability $P_{\mathrm{BS}}(\chi^2)$ is small for
the $D^+$ signal and is large for combinatorial background. 
Background distributions are taken from sidebands in the \kkpi mass, while 
signal distributions are obtained from the signal
regions, with the normalized sideband distributions subtracted.

\begin{figure}
  \vskip -0.15in
  \centering
  \includegraphics[width=\columnwidth]{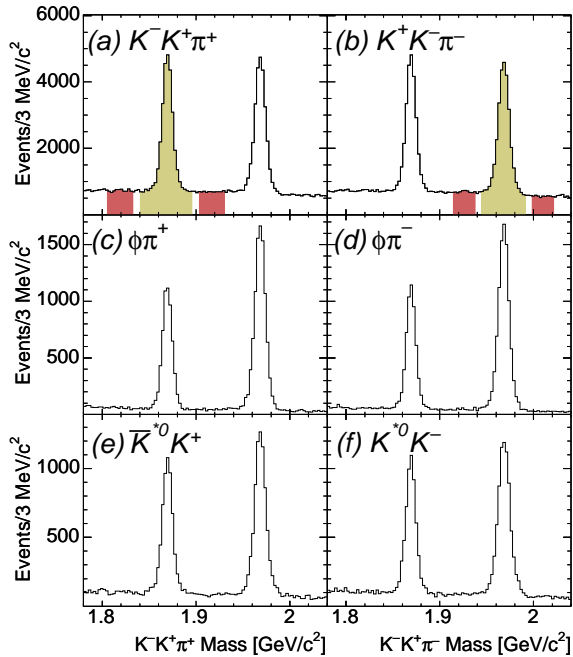}
  \caption{\label{fig:Yields}
    $KK\pi$ mass distributions for positively charged (left) and negatively
    charged (right) $D$ and $D_s$ candidates for events satisfying the requirement 
    $r \geq 4.3$.  Figures (a) and (b) are for all $KK\pi$ candidates, while
    (c) and (d) are for $\phi \pi$ candidates, and (e) and (f) for
    $\overline{K}^{*0} K$ candidates. Signal (yellow or light shaded) 
    and sidebands (red or darker
    shaded) regions are shown for \Dp and \Ds decays in (a) and (b),
    respectively.
  }
\end{figure}

For \Dpm decays, the $m_{KK\pi}$ signal band is defined as 
$[1.840, 1.896]$ \gevcc and the sideband mass regions as
$[1.805, 1.833]$ \gevcc and $[1.903, 1.931]$ \gevcc [see
Fig.~\ref{fig:Yields} (a)]. Product likelihoods are constructed
for the signal, $\mathcal{L}_{\mathrm{sig}} =\prod_{i}
\mathcal{L}^{i}_{\mathrm{sig}}(x_i)$, and the background,
$\mathcal{L}_{\mathrm{bkg}} =\prod_{i}
\mathcal{L}^{i}_{\mathrm{bkg}}(x_i)$, where $i$ runs over two or more of
the variables described. 

About 16\% of the events have more than one \Dp
meson candidate. For such events 
the candidate with the highest likelihood ratio is
selected.

The sensitivity $S/\Delta S$,
where $S$ and $\Delta S$ refer to the signal yield and 
its uncertainty, is optimized as a function of the product likelihood
ratio $r \equiv \mathcal{L}_{\mathrm{sig}} / \mathcal{L}_{\mathrm{bkg}}$ 
formed using $p_{\mathrm{CM}}$ and $P_{\mathrm{BS}}(\chi^2)$; the
optimal selection is found to be $r \geq 4.3$. 
This criterion is applied to both CF and SCS decays.
When we use the analogous ratio $r_1$ obtained by including the
PDF for $d_{xy}$ in $\mathcal{L}_{\mathrm{sig}}$ and
$\mathcal{L}_{\mathrm{bkg}}$, the sensitivity is nearly as
good. The results we find using $r_1$ provide a measure of systematic
uncertainty.

The subsamples for the decays \Dptophipi and
\Dptokstark are selected by requiring that the invariant mass of the
resonant decays be within 10~\mevcc and 50~\mevcc of the nominal
$\phi$ and \Kstarzb masses, respectively~\cite{pdg}. In addition, the  
resonant signal samples are enhanced 
by a selection on the cosine of the helicity angle ($\cos
\theta_H$).  In the \Dptophipi decay mode, the helicity angle is defined
as the angle between the $K^{-}$ and the $\pi^{+}$ in the $\phi$ rest
frame. In the \Dptokstark decay mode, the helicity angle is defined as
the angle between the $K^{-}$  
and the
$K^{+}$ in the $\overline{K}^{*0}$ rest frame. Maximum sensitivity is
obtained when $| \cos \theta_H | \geq 0.2$ and $| \cos \theta_H | \geq
0.3$ for \Dptophipi and for \Dptokstark, respectively.

The CF \Dsptokkpi decays are selected by a procedure identical to that
for the SCS \Dptokkpi decays.
We choose the signal $m_{KK\pi}$ region  to be $[1.944,1.992]$~\gevcc, 
while the sidebands are chosen to be $[1.914, 1.938]$ and
$[1.998, 2.022]$~\gevcc, respectively 
[see Fig.~\ref{fig:Yields} (b)]. In addition, 
contamination from \Dptokpipi decays is removed as follows: for
all $KK\pi$ candidates, the kaon with the same charge as the pion is
treated as a pion and then the $K\pi\pi$ invariant mass is
calculated. We observe a \Dp peak, indicating that part of the \Ds
signal is composed of 
misidentified \Dp candidates. Events in the region $1.855\leq
m_{K\pi\pi} \leq 1.883$~\gevcc are removed from the \Ds sample.
Contamination
from $D^{*+}\rightarrow D^0(\rightarrow
K^-\pi^+,K^-K^+)\pi^+$ decays is removed by eliminating events for which
$m_{K^-h^+}\geq 1.84$~\gevcc.
Candidates for \Dptokpipi are eliminated if both $K\pi$ combinations
satisfy the requirement. Partially reconstructed $D^{*+}\rightarrow
D^0(\rightarrow K^-\pi^+\pi^0)\pi^+$ decays can also be misidentified as
\kkpi candidates if the $\pi^0$ is missed and the charged pion is
misidentified as a kaon. Most of these decays are eliminated by
assigning a pion mass to kaon tracks and removing candidates for which the
mass difference $(m_{K^-\pi^+\pi^+} - m_{K^-\pi^+})$ lies in the range
$[0.139, 0.150]$~\gevcc. 

Figure~\ref{fig:Yields} shows the mass
distributions obtained after all selection criteria are applied. 
The yields, listed in Table~\ref{tbl:yields}, are
computed by subtracting from the number of events in the signal                                                         
region a scaled background estimate, obtained from the
sideband mass region. 

\begin{table}[htb]
  \caption{\label{tbl:yields}
    Yields of background subtracted events, separately for each charge.
  }
  \begin{center}
    \begin{tabular}{lcc}
      \hline\hline
      \multicolumn{1}{c}
      {Parent Charge} &     $+$       &         $-$      \\
      \hline

      \Dpmtokkpi   & $ 21632 \pm 228$ & $ 20940 \pm 226$ \\
      \Dpmtophipi  & $  5452 \pm  87$ & $  5327 \pm  86$ \\
      \Dpmtokstark & $  5247 \pm  96$ & $  5113 \pm  96$ \\
      \Dspmtokkpi  & $ 23066 \pm 217$ & $ 22928 \pm 214$ \\

      \hline\hline
    \end{tabular}
  \end{center}
\end{table}

The efficiencies needed for the $A_{CP}$ calculation are obtained from a
sample of MC generated $c\overline{c}$ events to which the same
selection criteria are applied.
The efficiencies for each
decay mode are shown in Table~\ref{tbl:Efficiencies}.

We obtain $A_{CP}$ using Eq.~(\ref{eqacp}) and replacing branching
fractions with efficiency-corrected yields. 
The results are shown in Table~\ref{tbl:Acp}.
We also studied the \CP asymmetry in 16 bins of the \Dptokkpi Dalitz plot
and found that the asymmetry is consistent with being constant (with a
probability of 51\%) and zero.

\begin{figure}
  \vskip -0.15in
  \centering
  \includegraphics[width=0.75\columnwidth]{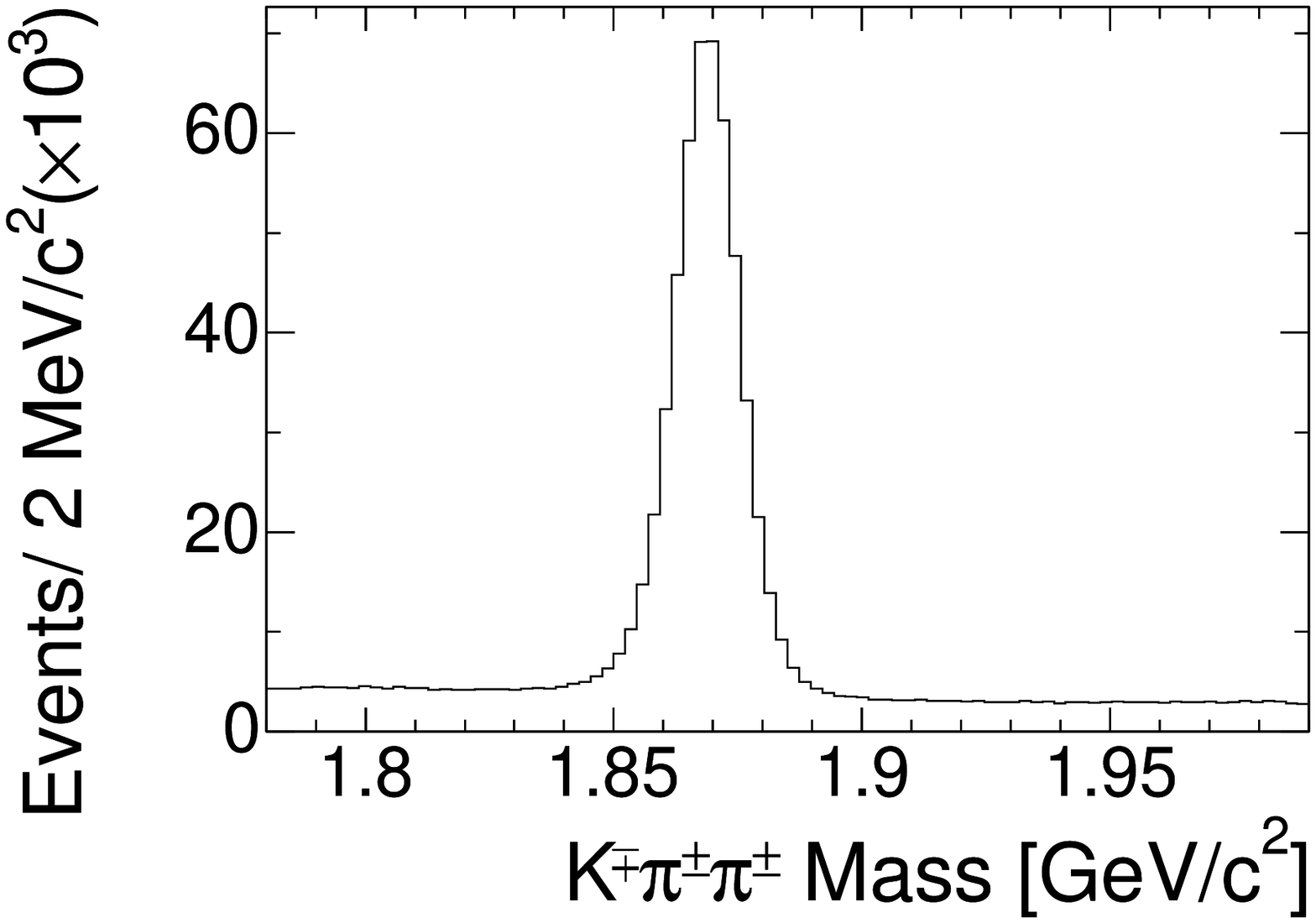}
  \caption{\label{fig:KpipiYield}
    Mass distribution for \Dpmtokpipi decays.
  }
\end{figure}

We use the CF sample of $\Dptokpipi$ decays, obtained using selection criteria
identical to the SCS case, to determine the relative branching fraction  
$\frac{\Gamma (\Dptokkpi )}{\Gamma(\Dptokpipi)}$ as
follows. The CF and SCS Dalitz plots are first divided into equally
populated bins (16 bins for the SCS mode, 64 for the CF mode). Next, the
signal and normalization yields and 
efficiencies are calculated bin-by-bin. The efficiency-corrected yields
are then summed and divided to obtain the
ratio. Figure~\ref{fig:KpipiYield} shows the mass distribution in the CF
$\Dpmtokpipi$ mode, for which the average efficiency is $10.03 \pm 0.01
(\textrm{stat.})$\%.  We obtain a relative branching fraction of
$(10.7\pm0.1 (\textrm{stat.}))\times 10^{-2}$.

\begin{table}[htb]
  \caption{\label{tbl:Efficiencies}
    Efficiencies for positively ($\varepsilon^+$) and
    negatively ($\varepsilon^-$) charged $D$ and $D_s$ meson
    decays. Efficiencies are in percent. 
    The stated uncertainties are due to MC statistics only.
  }
  \begin{center}
    \begin{tabular}{lr@{$\pm$}lr@{$\pm$}l}
      \hline\hline
      \multicolumn{1}{c}{Decay} & \multicolumn{2}{c}{$\varepsilon^+$} & \multicolumn{2}{c}{$\varepsilon^-$} \\
      \hline

      \Dpmtokkpi   & 8.20&0.04 & 8.26&0.04 \\
      \Dpmtophipi  & 7.67&0.07 & 7.63&0.07 \\
      \Dpmtokstark & 5.88&0.07 & 5.90&0.07 \\
      \Dspmtokkpi  & 3.77&0.02 & 3.79&0.02 \\

      \hline\hline
    \end{tabular}
  \end{center}
\end{table}

\begin{table}[htb]
  \caption{\label{tbl:Acp}
    Results of the \CP-asymmetry measurements, $A_{CP}$. Also listed are
    the values for $A^{(2)}_{CP}$, the asymmetry computed without the
    normalization mode.
  }
  \begin{center}
    \begin{tabular}{lcc}
      \hline\hline
      \multicolumn{1}{c}{Decay} & $A_{CP}~[10^{-2}]$ & $A^{(2)}_{CP}~[10^{-2}]$ \\
      \hline

      $K^-K^+\pi^{\pm}$  & $+1.36\pm1.01$ & $+2.07\pm0.84$ \\
      $\phi \pi^{\pm}$   & $+0.24\pm1.45$ & $+0.94\pm1.33$ \\
      $\Kstarzb K^{\pm}$ & $+0.88\pm1.67$ & $+1.58\pm1.57$ \\

      \hline\hline
    \end{tabular}
  \end{center}
\end{table}

\section{SYSTEMATIC UNCERTAINTIES AND CROSS-CHECKS}

The only difference between the final states from $D_s^\pm$ and $D^\pm$
decays considered here is a slightly harder momentum spectrum for the
$D_s^\pm$ decay products. In turn, these small differences are corrected
for by the efficiencies which come from MC. 
Any charge asymmetry in the detection of pions thus cancels
when \Dspmtokkpi decays are used as normalization, as in
Eq.~(\ref{eqacp}). We estimate the systematic uncertainty on the
\CP asymmetries by combining estimates of the contributions from various
identified sources listed in Table~\ref{tbl:Syst_Acp}.

The uncertainty 
due to small differences in momentum spectra of
$\pi$, $K$ from \Dp and \Ds decays, $0.06$\%, is estimated as three
times the maximum difference in $\pi$, $K$ asymmetries in
\Dp and \Ds decays. We evaluate an uncertainty for the background subtraction
by changing the widths of the sideband mass regions. The uncertainty is taken
to be the difference in the central values of $A_{CP}$. The uncertainties in
the likelihood-ratio technique are estimated with two variants: (i)
tightening the likelihood ratio to produce a $10$\% change in the 
yields, and (ii) using the likelihood ratio $r_1$ (described above) in
place of $r$. The systematic uncertainty is chosen to be the larger of the two
changes.
Table~\ref{tbl:Syst_Acp} summarizes these systematic uncertainties for
the observed \CP asymmetries.

\begin{table}[htb]
  \caption{
    \label{tbl:Syst_Acp}
    Systematic uncertainties for the \CP asymmetries.
  }
  \begin{center}
    \begin{tabular}{lccc}
      \hline\hline
       \vspace{-0.2cm} & & & \\
      \multicolumn{1}{c}{Source} & $K^-K^+\pi^{\pm}$ & $\phi \pi^{\pm}$ & $\Kstarzb K^{\pm}$ \\
      \multicolumn{1}{c}{} & $A_{CP}~[10^{-2}]$ & $A_{CP}~[10^{-2}]$ & $A_{CP}~[10^{-2}]$ \\
      \hline

      MC simulation        & $0.06$   & $0.06$   & $0.06$ \\
      Background estimate  & $0.63$   & $0.32$   & $0.49$ \\
      Selection criteria   & $0.46$   & $0.54$   & $0.54$ \\
      \hline
      Total                & $0.78$   & $0.63$   & $0.73$ \\
      \hline\hline

    \end{tabular}
    \end{center}
\end{table}

We performed two cross-checks on our measurement of $A_{CP}$. 
First, we calculated an alternative measure of \CP asymmetry without using
\Dsptokkpi decays as normalization, which we labeled $A^{(2)}_{CP}$ in
Table~\ref{tbl:Acp}. We find its values to be consistent with our
measurements of $A_{CP}$. Second, we measured the \CP asymmetry for a
control sample: the CF decays \Dsptokkpi (non-resonant as well as
resonant). This asymmetry is 
expected to be zero within the SM.  In \Dsptokkpi decays, both the
$D_s^+$ and the $D_s^-$ decay to two oppositely charged kaons and only
the pion charge differs in particle and antiparticle decays. Thus, any
detector-induced asymmetry would arise only from a charge asymmetry in
pion tracking and is expected to be very small. Indeed the measured
value is $(+0.6\pm0.8)\times 10^{-2}$.

As a final cross-check, the \CP asymmetry has also been studied as a
function of the \Dp laboratory momentum, as well as by the run period.
No significant dependence on momentum or detector operation conditions
is observed.

  A summary of the systematic uncertainties for the relative branching fraction 
$\frac{\Gamma (\Dptokkpi )}{\Gamma(\Dptokpipi)}$ is given in
Table~\ref{tbl:Syst_BR}. The fractional uncertainty due to PID and tracking
has been estimated as 2.1\%, computed as the sum
in quadrature of 1.1\%\ for PID and 1.8\%\ for
tracking~\cite{Ups4S}. The PID uncertainty is estimated from a comparison of
PID efficiencies in data and MC. The tracking uncertainty, which is the
uncertainty on the $K/\pi$ efficiency ratio, is conservatively estimated
as three times its value obtained using MC.

\begin{table}[htb]
  \caption{\label{tbl:Syst_BR}
    Systematic uncertainties for the relative branching fraction.
  }
  \begin{center}
    \begin{tabular}{lcc}
      \hline\hline
      \multicolumn{1}{c}{Source} & Uncertainty~$[10^{-2}]$ \\
      \hline

      PID + tracking         & $0.22$ \\
      Background estimate    & $0.05$ \\
      Selection criteria     & $0.02$ \\
      \hline
      Total                  & $0.23$ \\

      \hline\hline
    \end{tabular}
  \end{center}
\end{table}

\section{SUMMARY}

We have searched for a \CP asymmetry in \Dptokkpi,
\Dptophipi, and \Dptokstark decays and measured the relative branching fraction of
\Dptokkpi decays, with a data sample of $79.9$ \invfb collected by
the \babar\ experiment.

\begin{table}[htb]
  \caption{\label{tbl:AcpResults}
    Results of the \CP asymmetry ($A_{CP}$) measurements for \Dpm decays.
  }
  \begin{center}
    \begin{tabular}{lc}
      \hline\hline
      \multicolumn{1}{c}{Decay} & $A_{CP}~[10^{-2}]$ \\
      \hline

      $K^-K^+\pi^{\pm}$  & $+1.4 \pm 1.0 (\textrm{stat.}) \pm 0.8 (\textrm{syst.})$ \\
      $\phi \pi^{\pm}$   & $+0.2 \pm 1.5 (\textrm{stat.}) \pm 0.6 (\textrm{syst.})$ \\
      $\Kstarzb K^{\pm}$ & $+0.9 \pm 1.7 (\textrm{stat.}) \pm 0.7 (\textrm{syst.})$ \\

      \hline\hline
    \end{tabular}
  \end{center}
\end{table}
 
The measurements of the \CP asymmetries are summarized in 
Table~\ref{tbl:AcpResults}.  These
results are in agreement with previous published results~\cite{refcp},
with our results in the resonant modes having significantly smaller
uncertainties.

Further, we obtain a branching fraction for \Dptokkpi decays relative
to that for \Dptokpipi decays of
$(10.7\pm0.1 (\textrm{stat.})\pm0.2 (\textrm{syst.}))\times 10^{-2}$. This
result is a significant improvement over previous
measurements~\cite{refbr}.


\begin{acknowledgments}
We are grateful for the excellent luminosity and machine conditions
provided by our \pep2\ colleagues, 
and for the substantial dedicated effort from
the computing organizations that support \babar.
The collaborating institutions wish to thank 
SLAC for its support and kind hospitality. 
This work is supported by
DOE
and NSF (USA),
NSERC (Canada),
IHEP (China),
CEA and
CNRS-IN2P3
(France),
BMBF and DFG
(Germany),
INFN (Italy),
FOM (The Netherlands),
NFR (Norway),
MIST (Russia), and
PPARC (United Kingdom). 
Individuals have received support from CONACyT (Mexico), A.~P.~Sloan Foundation, 
Research Corporation,
and Alexander von Humboldt Foundation.

\end{acknowledgments}

\section{REFERENCES}


\begin{thebibliography}{99}
\bibitem{CCsign}
  Charge conjugation is assumed throughout the selection unless 
otherwise indicated.
\end{thebibliography}
\end{document}